\documentclass[11pt,a4paper]{JHEP3_mod2}

\usepackage{amsmath,latexsym,amssymb,slashed}
\usepackage{graphicx}				
\usepackage{amssymb}					
\usepackage{tabularx}				
\usepackage[vcentermath]{youngtab}	

\newcommand{\eq}[1]{\begin{equation}#1\end{equation}}
\newcommand{\spl}[1]{\begin{split}#1\end{split}}

\newcommand\ga{\gamma}

\newcommand{\boxedeq}[1]{
\begin{equation}
\fbox{
\rule[0.7cm]{0pt}{0pt}
$#1$
\rule[-0.45cm]{0pt}{0pt}
}
\end{equation}
}

\def\d{\text{d}}

\author{Robin Terrisse and Dimitrios Tsimpis\\
Institut de Physique Nucl\'{e}aire de Lyon\\
Universit\'{e} de Lyon, UCBL, 
UMR 5822, CNRS/IN2P3, \\
4 rue Enrico Fermi,
69622 Villeurbanne Cedex,  France\\

E-mail:
\email{terrisse@ipnl.in2p3.fr}, \email{tsimpis@ipnl.in2p3.fr}}
\abstract{
We construct a 
consistent four-scalar truncation of ten-dimensional IIA supergravity on nearly K\"{a}hler  spaces in the presence of dilatino condensates.  
The truncation is universal, i.e.~it does not depend on any detailed features of the compactification manifold other than its nearly K\"{a}hler property, and admits a 
smooth limit to a universal four-scalar consistent truncation on Calabi-Yau spaces. 
The theory admits formal solutions with nonvanishing condensates, of the form $S^{1,3}\times M_6$, where $M_6$ is a six-dimensional nearly K\"{a}hler or Calabi-Yau manifold, and $S^{1,3}$ 
can be de Sitter, Minkowski or anti-de Sitter four-dimensional space. 
}
\title{Consistent truncation with dilatino condensation on nearly K\"{a}hler and Calabi-Yau manifolds}
\preprint{}

\newcommand{\swed}{{\scriptscriptstyle \wedge}}

\begin{document}
\setlength{\parindent}{0pt}
\newpage



\section{Introduction}

Theories with fermionic condensates are interesting  in a variety of contexts, not least for cosmological applications. 
They have been studied in the past mainly in heterotic theory \cite{Dine:1985rz, Derendinger:1985kk, LopesCardoso:2003sp, Derendinger:2005ed, Manousselis:2005xa, Chatzistavrakidis:2012qb, Gemmer:2013ica, Quigley:2015jia, Minasian:2017eur} and  eleven-dimensional supergravity \cite{Duff:1982yi,Jasinschi:1986ze}. 
More recently the IIA theory has been shown to admit 
dilatino condensates which, in contrast to the situation in heterotic theory,  can lead to a positive cosmological term and (formal) de Sitter solutions \cite{Soueres:2017lmy}. 
In the present paper we show that  theories with fermionic condensates can also admit consistent truncations. To our knowledge this is the first time 
a consistent truncation (CT) of a higher-dimensional theory has been constructed in the presence of condensates.

 A CT of a higher-dimensional theory is a lower-dimensional theory all of whose solutions lift to solutions of the higher-dimensional theory. 
CT's have proven to be useful in various different contexts, notably in applications of AdS/CFT. 
Although a CT is not a prerequisite for performing the holographic analysis of a theory, the truncation reduces the higher-dimensional equations of motion
to simpler lower-dimensional equations, thus facilitating the construction of a precise holographic dictionary. 

CT's have mostly been constructed on homogeneous spaces such as spheres \cite{deWit:1986oxb,Nastase:1999cb,Nastase:1999kf,Lu:1999bw,Cvetic:2000nc,Guarino:2015jca}, 
often exploiting the left-invariant forms present in a coset description of the space \cite{MuellerHoissen:1987cq,Kapetanakis:1992hf,Bena:2010pr,Cassani:2011fu,Cassani:2012pj}. 
Recent reformulations of supergravity using double field theory and  exceptional generalized geometry have also proven to be powerful tools in the 
construction of CT's \cite{Aldazabal:2011nj,Geissbuhler:2011mx,Lee:2014mla,Hohm:2014qga,Cassani:2016ncu,Baguet:2015sma,Baguet:2015iou,Ciceri:2016dmd,Inverso:2016eet,Inverso:2017lrz,Malek:2017njj,Malek:2018zcz}. 
Another approach that has been used in the past consists  
in exploiting the $G$-structure of the internal space to guide the construction  of the truncation ansatz 
\cite{Gauntlett:2007ma,Cassani:2009ck,Cassani:2010uw,Skenderis:2010vz,Gauntlett:2010vu,Liu:2010sa}.

In the present paper  we will use the latter approach to construct a 
CT of ten-dimensional (massive) IIA supergravity on nearly K\"{a}hler (NK) spaces in the presence of dilatino condensates.  
Moreover, taking the limit of vanishing scalar curvature, we establish the consistency of a four-scalar truncation of 
IIA supergravity on Calabi-Yau (CY) manifolds. 
For vanishing dilatino condensate it should coincide with the CT of \cite{Guarino:2015vca}, cf.~footnote 5 therein. 
To our knowledge,  there are no other known  truncations on a CY manifold with 
massive lower-dimensional fields, whose consistency has been rigorously proven to date.

There are four homogeneous compact NK six-manifolds: $S^6$, $S^3\times S^3$, $\mathbb{CP}^3$ and $\mathbb{F}^{1,2}$ \cite{2006math.....12655B}. Until recently these were 
the only known compact NK six-manifolds.  Two nonhomogenous examples were constructed in \cite{Foscolo:2015vqa}, and it 
is expected that many more  should exist \cite{2010arXiv1011.4681S}. 
CT's on homogeneous NK manifolds have been constructed in the past, relying on the coset description of these spaces \cite{Cassani:2009ck}.\footnote{A reduction on NK spaces leading to four-dimensional $\mathcal{N}=2$  gauged supergravity, has been performed in \cite{KashaniPoor:2007tr}, 
but without a complete proof of consistency. In this  approach one postulates the existence  of a finite set of forms on the internal manifold satisfying a certain list of constraints 
 \cite{Gurrieri:2002wz, DAuria:2004kwe, House:2005yc, Grana:2005ny, Louis:2006kb, KashaniPoor:2006si}. Plugging the truncation ansatz into the higher-dimensional action can then be shown to 
 result in a lower-dimensional  gauged supergravity. The consistency of this procedure, albeit plausible, has not been 
 rigorously proven to date. See \cite{Andriot:2018tmb} for a recent discussion of this approach. In a subsequent development, the authors of \cite{Guarino:2015vca} 
 proved the consistency of \cite{KashaniPoor:2007tr} by showing that it coincides with the $G_2$-invariant subsector of the $\mathcal{N}=8$ ISO(7) dyonic supergravity arising from a consistent 
 truncation of IIA on $S^6$.}  
 The CT  we will present in this paper is ``universal'', meaning 
 it only relies on the NK property of the manifold, but not on any of its detailed features. 
In particular the CT is valid for any NK manifold whether or not it is homogeneous and admits a coset description.

The outline of the paper is as follows. In section \ref{sec:gen} we review the dilatino-condensate action. The solutions with nonvanishing condensates are 
presented in section \ref{sec:condensate}, after reviewing the bosonic AdS$_4$ solutions in section \ref{sec:ads}. Following a discussion of the case without condensates in section \ref{sec:massive}, the consistent truncation in the presence of condensates is given in \eqref{ctrcs}-\eqref{fgcgg} of section \ref{sec:3dil}. Its CY limit is presented in section \ref{sec:cylt}.  
 We conclude with a discussion in section \ref{sec:disc}.

\section{Dilatonic solutions}\label{sec:gen}

In a maximally-invariant vacuum of the theory, all fermion vacuum expectation values (VEV) are assumed to vanish, 
but quadratic or quartic fermion terms may still develop nonvanishing VEV's. Schematically,
\eq{\label{v1}
\langle \lambda\rangle=0~;~~~
\langle\bar{\lambda}\lambda\rangle:=\int [\mathcal{D}\Phi](\bar{\lambda}\lambda) e^{-S[\Phi]}\neq 0~,
}
where $\lambda$ collectively denotes the fermions and $\Phi$ stands  for all fields in the action $S[\Phi]$. 
The vacuum $\langle\Phi\rangle$ is obtained by minimizing the effective action $S_{\text{eff}}$ with respect to the fields,
\eq{
\left.\frac{\delta S_{\text{eff}}}{\delta\Phi}\right|_{\langle\Phi\rangle}\!\!\!\!=0
~,}
where, at tree level in the coupling, the effective action coincides with $S$. Moreover, in the case of the IIA superstring, the two-derivative effective action $S_{\text{eff}}$ coincides with the action of ten-dimensional IIA supergravity to all orders in 
string perturbation.\footnote{Loop corrections in the string coupling are expected to modify the terms in $S_{\text{eff}}$ with eight or more derivatives. In general, this will no longer be the case in the  compactified theory.}  
However, nonperturbatively,  $S_{\text{eff}}$ may develop nonvanishing VEV's for the quadratic and quartic fermion terms.

We will not examine the mechanism for the generation of fermionic condensates here: we will simply assume their presence  and examine the implications. 
In the following we will look in particular for {\it dilatonic solutions}, i.e.~for solutions of the {\it dilatino-condensate action} of \cite{Soueres:2017lmy}. This is obtained from the IIA supergravity
 action by setting the Einstein-frame gravitino to zero. Moreover, the quadratic and quartic dilatino terms in the action should be thought of as replaced by their 
 condensate VEV's, and thus become (constant) parameters of the action. 
The dilatino-condensate action  should therefore be regarded as a book-keeping device 
whose variation with respect to the bosonic fields gives the correct bosonic equations of motion in the presence of dilatino condensates; the fermion equations of motion are trivially satisfied in the maximally-invariant vacuum, and need not be considered.

In \cite{Soueres:2017lmy} the fermionic terms of IIA supergravity were determined in the ten-dimensional superspace formalism previously developed in \cite{Tsimpis:2005vu}, resolving an ambiguity in the original literature \cite{Giani:1984wc,Campbell:1984zc,Huq:1983im} concerning the quartic fermions, and finding agreement with \cite{Giani:1984wc}. 
In the conventions of \cite{Soueres:2017lmy}, the dilatino-condensate action of (massive) IIA reads,\footnote{We have rescaled the Romans mass: $m\rightarrow 5m/4$ with respect to \cite{Soueres:2017lmy}. Moreover $\hat{g}_{mn}$ 
of that reference is denoted ${g}_{mn}$ here. We have also changed conventions for the Riemann tensor so that $\hat{R}$ of \cite{Soueres:2017lmy} is $-R$ here.}  
\eq{\spl{\label{action3}
S=S_b+
\int\d^{10}x\sqrt{{g}} \Big\{ 
&(\bar{\Lambda}\Gamma^M\nabla_M\Lambda)
-\frac{21}{16}e^{5\phi/4}m(\bar{\Lambda}\Lambda)+\frac{3}{512}(\bar{\Lambda}\Lambda)^2\\
&-\frac{5}{32}e^{3\phi/4} F_{MN}(\bar{\Lambda}\Gamma^{MN}\Gamma_{11}\Lambda)
+\frac{1}{128}e^{\phi/4} G_{MNPQ}(\bar{\Lambda}\Gamma^{MNPQ}\Lambda)
\Big\}
~,}}
where $\Lambda$ is the dilatino; $S_b$ is the bosonic sector  of Romans supergravity \cite{Romans:1985tz},  
\eq{\spl{\label{ba}S_b=\int\d^{10}x\sqrt{{g}}\Big(
&-{R}+\frac12 (\partial\phi)^2+\frac{1}{2\cdot 2!}e^{3\phi/2}F^2\\
&+\frac{1}{2\cdot 3!}e^{-\phi}H^2+\frac{1}{2\cdot 4!}e^{\phi/2}G^2
+\frac{1}{2}m^2e^{5\phi/2}\Big) 
+\mathrm{CS}
~,
}}
where  CS denotes the Chern-Simons term. 
We emphasize that, as mentioned previously, the dilatino terms in \eqref{action3} are not dynamical but should be thought of as parameters of the action. In particular 
$(\bar{\Lambda}\Lambda)^2$ should be thought of as the VEV $\langle(\bar{\Lambda}\Lambda)^2\rangle$ and is therefore a priori independent of $(\bar{\Lambda}\Lambda)$, which 
should be thought of as the VEV $\langle\bar{\Lambda}\Lambda\rangle$.

The dilaton and Einstein equations following from action (\ref{action3}) read,
\eq{\spl{\label{beomf1}
0&=-{\nabla}^2\phi+\frac{3}{8}e^{3\phi/2}F^2-\frac{1}{12}e^{-\phi}H^2+\frac{1}{96}e^{\phi/2}G^2 +\frac{5}{4}m^2e^{5\phi/2}\\
&-\frac{105}{64}e^{5\phi/4}m(\bar{\Lambda}\Lambda)
-\frac{15}{128}e^{3\phi/4} F_{MN}(\bar{\Lambda}\Gamma^{MN}\Gamma_{11}\Lambda)
+\frac{1}{512} e^{\phi/4} G_{MNPQ}(\bar{\Lambda}\Gamma^{MNPQ}\Lambda)
~,
}}
and,
\eq{\spl{\label{beomf2}
{R}_{MN}&=\frac{1}{2}\partial_M\phi\partial_N\phi+\frac{1}{16}m^2e^{5\phi/2}{g}_{MN}
+\frac{1}{4}e^{3\phi/2}\Big(  2F^2_{MN} -\frac{1}{8} {g}_{MN}  F^2 \Big)\\
&+\frac{1}{12}e^{-\phi}\Big(  3H^2_{MN} -\frac{1}{4} {g}_{MN}  H^2 \Big)
+\frac{1}{48}e^{\phi/2}\Big(   4G^2_{MN} -\frac{3}{8} {g}_{MN}  G^2 \Big)\\
&+
\frac12(\bar{\Lambda}\Gamma_{(M}\nabla_{N)}\Lambda)
+\frac{1}{16}g_{MN}(\bar{\Lambda}\Gamma^{P}\nabla_{P}\Lambda)
-\frac18 {g}_{MN}\Big[ \frac{21}{16}e^{5\phi/4}m(\bar{\Lambda}\Lambda)
-\frac{3}{512}(\bar{\Lambda}\Lambda)^2
\Big]\\
&-\frac{5}{32}e^{3\phi/4} 
F_{(M}{}^P(\bar{\Lambda}\Gamma_{N)P}\Gamma_{11}\Lambda)
+\frac{1}{128}e^{\phi/4} 
\Big[
2G_{(M}{}^{PQR}(\bar{\Lambda}\Gamma_{N)PQR}\Lambda)
-\frac18 {g}_{MN}
G_{(4)}(\bar{\Lambda}\Gamma^{(4)}\Lambda)
\Big]
~.
}}
The form equations read,
\eq{\spl{\label{beomf3}
0&=\d {\star}\big[ e^{3\phi/2}F  
-\frac{5}{16}e^{3\phi/4} (\bar{\Lambda}\Gamma_{(2)}\Gamma_{11}\Lambda)
\big]+e^{\phi/2}H\swed {\star} G\\
0&=\d{\star} 
e^{-\phi}H
+e^{\phi/2}F\swed{\star} G-\frac{1}{2}G\swed G+m e^{3\phi/2}{\star} F\\
0&=\d
{\star} 
\big[
e^{\phi/2}G
+\frac{3}{16}e^{\phi/4} (\bar{\Lambda}\Gamma_{(4)}\Lambda)
\big]
-H\swed G
~,
}}
where: 
$(\bar{\Lambda}\Gamma_{(p)}\Lambda):=\frac{1}{p!} (\bar{\Lambda}\Gamma_{M_1\dots M_p}\Lambda)\d x^{M_p}\wedge\dots\wedge\d x^{M_1}$. Moreover the forms obey the following Bianchi identities,
\eq{\label{bi}
\d F= m H~;~~~\d H=0~;~~~\d G=H\wedge F
~.}

\subsection{Bosonic AdS$_4$ solutions}\label{sec:ads}

The equations of motion \eqref{beomf1}-\eqref{bi} admit bosonic solutions of the form 
AdS$_4\times M_6$, where $M_6$ is nearly  K\"{a}hler, cf.~section 11.4 of \cite{Lust:2008zd}. Let us now review these 
solutions, before switching on the  dilatino condensates in section \ref{sec:condensate}.

We take the ten-dimensional spacetime to be of direct product form AdS$_4\times M_6$, 
\eq{\label{dpst}\d s^2=\d s^2(\text{AdS}_4)+ \d s^2(M_{6})
~.}
Let us parameterize,
\eq{\label{ei43}
{R}_{\mu\nu}= 3\Omega ~\!g_{\mu\nu}~;~~~
{R}_{mn}= 20\omega^2 g_{mn}
 ~,}
where  $g_{\mu\nu}$,   $g_{mn}$ are the components of the metric in the 
external, internal space respectively;  $\Omega$  is negative for  anti-de Sitter space; $\omega$ is related to the first 
torsion class of $M_6$ through (\ref{torsionclassesbnk}).

Moreover we set the dilaton  to zero, $\phi=0$,  and we parameterize the three-form and RR fluxes as follows,
\eq{\label{ma}H=f\mathrm{Re}\Omega~;~~~
F=b J~;~~~G=a~\!\mathrm{vol}_4+\frac12 c J^2~;~~~f,a,b,c\in\mathbb{R}
~,}
where $J$ is the K\"{a}hler form of $M_6$, and $\mathrm{vol}_4$ is the volume element of AdS$_4$.  
It is then straightforward to 
see, using (\ref{torsionclassesbnk}),  that the Bianchi identities (\ref{bi}) are satisfied provided,
\eq{\label{hbi}
mf+6b\omega=0
~.}
The $F$-form equation in (\ref{beomf3}) is automatically satisfied, while the $H$-form equation reduces to,
\eq{\label{e21}
2bc-  ac+  mb-8f\omega=0~.}
The $G$-form equation in (\ref{beomf3}) reduces to,
\eq{\label{e215}
a f+6c\omega=0~.}
Moreover the dilaton equation reduces to,
\eq{\label{e22}
0=9b^2+3c^2+ 5m^2-a^2-8f^2~.}
The mixed $(\mu,m)$  components of the Einstein equations are automatically satisfied, 
while 
the internal  $(m,n)$  components 
of the Einstein equations reduce to,
\eq{\label{e24}
20\omega^2= 2b^2+c^2 +m^2-f^2
~,}
where we have taken (\ref{e22}) into account. 
Finally the $(\mu,\nu)$ components of the Einstein equations reduce to,
\eq{\label{e23}
\Omega=\frac16f^2-10\omega^2 
~,}
where we have used (\ref{e22}) and (\ref{e24}).

As noted in \cite{Lust:2008zd}, these equations admit three general classes of solutions only one of which is supersymmetric 
and corresponds to the NK solutions first discovered in \cite{Behrndt:2004km}; it reads,
\eq{\label{sf}
a^2=\frac{27}{5} m^2~;~~~b=\frac19 a~;~~~c=\frac35 m~;~~~f=\pm\frac25 m~;~~~\omega=-\frac19 a~;~~~\Omega=-\frac{16}{25}m^2
~.}
In particular we see that the solution is parameterized by a single parameter (the Romans mass) and reduces to flat space without flux in the massless limit $m\rightarrow 0$.

\subsection{Solutions with dilatino condensates}\label{sec:condensate}

We will now allow for nonvanishing dilatino bilinear and quadratic condensates. 
Let $\Lambda_{\pm}$ be the positive-, negative-chirality components of the ten-dimensional dilatino. We decompose, 
\eq{
\Lambda_+=\theta_+\otimes\eta-\theta_-\otimes\eta^c~; ~~~\Lambda_-=\theta'_+\otimes\eta^c-\theta'_-\otimes\eta~,
}
where $\theta_+$, $\theta'_+$ are arbitrary anticommuting four-dimensional Weyl spinors of positive chirality, see appendix \ref{sec:conventions} for our spinor conventions.  
Furthermore the reality of $\Lambda_{\pm}$ imposes the conditions,
\eq{\bar{\theta}_+=\widetilde{\theta}_-
~; ~~~ \bar{\theta}_-=-\widetilde{\theta}_+
~,}
which implies in particular,
\eq{(\widetilde{\theta}_+{\theta}'_+)^* =-(\widetilde{\theta}_-\widetilde{\theta}'_-) 
~.}
We define the following three complex numbers parameterizing the four-dimensional dilatonic condensate,
\eq{
\mathcal{A}:=(\widetilde{\theta}_+{\theta}'_+)~;~~~
\mathcal{B}:=(\widetilde{\theta}_+{\theta}_+)~;~~~
\mathcal{C}:=(\widetilde{\theta}'_+{\theta}'_+)~.
}
In terms of these, the ten-dimensional dilaton bilinears decompose as follows,
\eq{\spl{
(\bar{\Lambda}_+ \Lambda_-)&=2\text{Re}(\mathcal{A})\\
(\bar{\Lambda}_+\Gamma_{mn} \Lambda_-)&=2\text{Im}(\mathcal{A})J_{mn}\\
(\bar{\Lambda}_+\Gamma_{mnrs} \Lambda_-)&=-6\text{Re}(\mathcal{A})J_{[mn}J_{rs]}\\
(\bar{\Lambda}_+\Gamma_{mnp} \Lambda_+)&=2\text{Re}(\mathcal{B}\Omega_{mnp})\\
(\bar{\Lambda}_-\Gamma_{mnp} \Lambda_-)&=-2\text{Re}(\mathcal{C}\Omega^*_{mnp})\\
(\bar{\Lambda}_+\Gamma_{\mu\nu\rho\sigma} \Lambda_-)&=2\text{Im}(\mathcal{A})\varepsilon_{\mu\nu\rho\sigma}
~,}}
where we have used the formul\ae{} of appendix \ref{app:a}. 
For the ``kinetic'' bilinear terms we will assume that $(\bar{\Lambda}_{\pm}\Gamma_{\mu}\nabla_{\nu} \Lambda_{\pm})=0$. 
 For a NK internal manifold 
such that (\ref{torsionclassesnk}) holds, we have,
\eq{\spl{
(\bar{\Lambda}_+\Gamma_{(m}\nabla_{n)} \Lambda_+)&=-2\text{Im}(\mathcal{B}\omega)g_{mn}\\
(\bar{\Lambda}_-\Gamma_{(m}\nabla_{n)} \Lambda_-)&=-2\text{Im}(\mathcal{C}\omega^*)g_{mn}
~.}}
Let us now substitute the above in the 10d equations of motion, while retaining the same form  ansatz (\ref{ma}) as in section \ref{sec:ads}. The only difference is that we 
postulate a 10d metric of the form
\eq{\label{dpst}\d s^2=\d s^2({S}^{1,3})+ \d s^2(M_{6})
~,}
where now ${S}^{1,3}$ can be any maximally symmetric four-dimensional space.  
I.e.~(\ref{ei43}) is still valid here, but we allow $\Omega$ to also be positive or zero (corresponding to de Sitter or Minsowski) in addition to anti-de Sitter.

With this ansatz the 10d equations of motion  are modified as follows: 
 the Bianchi identities (\ref{bi}) are satisfied provided,
\eq{\label{hbidc}
mf+6b\omega=0
~,}
as was the case for vanishing condensate. 
The $F$-form equation in (\ref{beomf3}) is automatically satisfied, while the $H$-form equation reduces to,
\eq{\label{e21dc}
2bc-  ac+  mb-8f\omega=0~,}
exactly as in the case of vanishing condensate. 
The $G$-form equation in (\ref{beomf3}) reduces to,
\eq{\label{e215dc}
(a +\frac34\text{Im}\mathcal{A})f+6(c-\frac34\text{Re}\mathcal{A})\omega=0~,}
thus receiving a contribution from the condensate. 
The dilaton equation reduces to,
\eq{\label{e22cd}
0=9b(b+\frac54\text{Im}\mathcal{A})+3c(c-\frac34\text{Re}\mathcal{A})+ 5m(m-\frac{21}{4}\text{Re}\mathcal{A})-a(a+\frac34\text{Im}\mathcal{A})-8f^2~.
}
The mixed $(\mu,m)$  components of the Einstein equations are automatically satisfied as before, 
while 
the internal  $(m,n)$  components 
of the Einstein equations reduce to,
\eq{ \spl{\label{e24cd}
20\omega^2 &= \frac{1}{16}m^2
+\frac{5}{16}b^2
+\frac12 f^2+\frac{7}{16}c^2+\frac{3}{16}a^2
\\
&+\frac58 b\text{Im}\mathcal{A}+\frac{3}{32}a\text{Im}\mathcal{A}
-\frac{15}{32}c\text{Re}\mathcal{A}-\frac74\text{Im}(\mathcal{B}\omega+\mathcal{C}\omega^*)
-\frac{21}{32} 
m\text{Re}\mathcal{A}+\frac{3}{2^{12}}(\bar{\Lambda} \Lambda)^2
~.}}
As already mentioned,  the last term above should be thought of as the VEV of a quartic fermion  term, thus 
a priori different from the square of the bilinear VEV.
Finally the $(\mu,\nu)$ components of the Einstein equations reduce to,
\eq{ \spl{\label{e23cd}
\Omega &= \frac{1}{48}m^2
-\frac{1}{16}b^2
-\frac16 f^2-\frac{3}{16}c^2-\frac{5}{48}a^2
\\
&-\frac{3}{32}a\text{Im}\mathcal{A}
+\frac{3}{32}c\text{Re}\mathcal{A}-\frac14\text{Im}(\mathcal{B}\omega+\mathcal{C}\omega^*)
-\frac{7}{32} 
m\text{Re}\mathcal{A}+\frac{1}{2^{12}}(\bar{\Lambda} \Lambda)^2
~.}}
In the limit of vanishing condensates one recovers the bosonic AdS$_4\times M_6$ solutions reviewed in section \ref{sec:ads}. Moreover one can obtain $\Omega>0$, and thus 
four-dimensional de Sitter space, e.g.~for $(\bar{\Lambda} \Lambda)^2$ sufficiently large.

\section{Consistent truncation}\label{sec:massive}

The solutions of section \ref{sec:gen} can be recovered from the equations of motion of a four-dimensional consistent truncation of the 
ten-dimensional IIA dilatino-condensate action. In the following we will construct the consistent truncation in the case of vanishing condensates. 
The case of nonvanishing condensates will be considered in section \ref{sec:3dil}.

Our ansatz for the ten-dimensional metric includes two new scalars $A$, $B$ with four-dimensional spacetime dependence,
\eq{\label{tdma}\d s^2_{(10)} =e^{2A(x)}\left(e^{2B(x)} g_{\mu\nu}\d x^{\mu}\d x^{\nu}+g_{mn}\d y^m\d y^n 
\right)~.
}
From the above we obtain the following formula for the ten-dimensional Laplacian of a scalar $S(x)$ with only four-dimensional spacetime dependence,
\eq{\nabla^2_{(10)} S(x)
=e^{-2A-2B}\left(  \nabla^2_{(4)} S(x)+ 
8\partial^{\rho}A\partial_{\rho}S+2\partial^{\rho}B\partial_{\rho}S
\right)
~,}
where the contractions on the right-hand side are taken with respect to the unwarped four-dimensional metric.
The Einstein tensor of  (\ref{tdma}) reads,
\eq{\spl{
R^{(10)}_{mn}&=R^{(6)}_{mn}-e^{-2B}g_{mn}\left(\nabla^{\rho}\partial_{\rho}A+
8\partial^{\rho}A\partial_{\rho}A+2\partial^{\rho}A\partial_{\rho}B
\right)\\
R^{(10)}_{\mu\nu}&=R^{(4)}_{\mu\nu}
-g_{\mu\nu}\left(\nabla^{\rho}\partial_{\rho}A+\nabla^{\rho}\partial_{\rho}B+
8\partial^{\rho}A\partial_{\rho}A+2\partial^{\rho}B\partial_{\rho}B+10\partial^{\rho}A\partial_{\rho}B\right)
\\
&+8\partial_{\mu}A\partial_{\nu}A+2\partial_{\mu}B\partial_{\nu}B
+16\partial_{(\mu}A\partial_{\nu)}B-8\nabla_{\mu}\partial_{\nu}A-2\nabla_{\mu}\partial_{\nu}B
~,}}
while the mixed components $R^{(10)}_{m\mu}$ vanish identically.

Our ansatz for the forms is such that the Bianchi identities (\ref{bi}) are automatically satisfied. It is given in terms of three scalars  $\varphi$, $\chi$, $\gamma$ which 
are taken to only carry four-dimensional spacetime dependence. Explicitly,
\eq{\label{forans}
F=m\chi J~;~~~ H=\d\chi \swed J-6\omega \chi \text{Re}\Omega~;~~~
G=\varphi\text{vol}_4+\frac12 (m\chi^2 +\gamma)J\swed J-\frac{1}{8\omega}\d\gamma\swed\text{Im}\Omega
~.}
%
%
%
%
In particular we obtain,
\eq{\spl{
F^2_{mn}&= m^2\chi^2e^{-2A}g_{mn}~;~~~F^2= 6m^2\chi^2e^{-4A}\\
H^2_{mn}&=2e^{-4A-2B}(\partial\chi)^2g_{mn} +144e^{-4A}\omega^2\chi^2 g_{mn}~;~~~
H^2_{\mu\nu}=6e^{-4A}\partial_{\mu}\chi\partial_{\nu}\chi\\
H^2&=18e^{-6A-2B}(\partial\chi)^2+864e^{-6A}\omega^2\chi^2\\
G^2_{mn}&=12e^{-6A} \left( m\chi^2+\gamma \right)^2 g_{mn}+\frac{3}{16\omega^2}e^{-6A-2B}(\partial\gamma)^2g_{mn}\\
G^2_{\mu\nu}&=-6e^{-6A-6B}\varphi^2g_{\mu\nu}  +\frac{3}{8\omega^2}e^{-6A} \partial_{\mu}\gamma\partial_{\nu}\gamma\\
G^2&= -24 e^{-8A-8B}\varphi^2+72 e^{-8A} \left( m\chi^2+\gamma \right)^2
+\frac{3}{2\omega^2}e^{-8A-2B}(\partial\gamma)^2
~,}}
where the contractions on the left-hand sides are taken with respect to the ten-dimensional metric while the contractions 
on the right-hand sides are taken with respect to the unwarped four- and six-dimensional metrics. The following expressions are also 
useful,
\eq{\spl{\label{hodsr}
\star_{10} F &=  \frac12 m\chi e^{6A+4B} \text{vol}_4\swed J^2\\
\star_{10} H &=  \frac12 e^{4A+2B} \star_{4}\!\d\chi\swed   J^2
+6\omega\chi e^{4A+4B}\text{vol}_4\swed \text{Im}\Omega
\\
\star_{10} G &= -\frac16 \varphi e^{2A-4B}  J^3+(m\chi^2+\gamma) e^{2A+4B} \text{vol}_4\swed J+\frac{1}{8\omega} e^{2A+2B}   \star_{4}\!\d\gamma\swed  \text{Re}\Omega
~,}}
where the Hodge star is defined as in \cite{Lust:2004ig, Soueres:2017lmy}. 
Plugging the above ansatz into the equations of motion we obtain the following: the internal $(m,n)$-components of the Einstein 
equations (\ref{beomf2}) read,
\eq{\spl{\label{et1}
0&=e^{-8A-2B}\nabla^{\mu}\left(
e^{8A+2B}\partial_{\mu}A
\right)+\frac{1}{16}m^2e^{5\phi/2+2A+2B}
+\frac{5}{16}e^{3\phi/2-2A+2B}m^2\chi^2\\
&+\frac18e^{-\phi-4A}(\partial\chi)^2
+18 e^{-\phi-4A+2B}\omega^2\chi^2
+\frac{1}{16}e^{\phi/2}
\left(
3e^{-6A-6B}\varphi^2+7e^{-6A+2B}( m\chi^2+\gamma)^2
\right)\\
&+\frac{1}{256\omega^2}e^{\phi/2-6A}(\partial\gamma)^2
-20e^{2B}\omega^2
~,}}
where we have taken (\ref{ei43}) into account. The external $(\mu,\nu)$-components read,
\eq{\spl{\label{et2}
R^{(4)}_{\mu\nu}&=
g_{\mu\nu}\left(\nabla^{2}A+\nabla^{2} B+
8(\partial A)^2+2(\partial B)^2+10\partial A\cdot \partial B\right)
\\
&-8\partial_{\mu}A\partial_{\nu}A-2\partial_{\mu}B\partial_{\nu}B
-16\partial_{(\mu}A\partial_{\nu)}B+8\nabla_{\mu}\partial_{\nu}A+2\nabla_{\mu}\partial_{\nu}B\\
&+\frac32e^{-\phi-4A} 
\partial_{\mu}\chi\partial_{\nu}\chi
+\frac12\partial_{\mu}\phi\partial_{\nu}\phi
+\frac{1}{32\omega^2}e^{\phi/2-6A}\partial_{\mu}\gamma\partial_{\nu}\gamma
 \\
&+\frac{1}{16}g_{\mu\nu}\Big(  
-\frac{3}{16\omega^2}e^{\phi/2-6A}(\partial\gamma)^2
-6e^{-\phi-4A}(\partial\chi)^2
\\
&+ m^2e^{5\phi/2+2A+2B}
-3m^2\chi^2 e^{3\phi/2-2A+2B} 
-288e^{-\phi-4A+2B}\omega^2\chi^2\\
&-5e^{\phi/2-6A-6B}\varphi^2 
-9e^{\phi/2-6A+2B}( m\chi^2+\gamma)^2 
\Big)
~,}}
while the mixed $(\mu,m)$-components are automatically satisfied. 
The dilaton equation reads,
\eq{\spl{\label{et3}
0&=e^{-10A-4B}\nabla^{\mu}\left(
e^{8A+2B}\partial_{\mu}\phi
\right)-\frac{5}{4}m^2e^{5\phi/2}
-\frac{9}{4}e^{3\phi/2-4A}m^2\chi^2
-\frac{1}{64\omega^2}e^{\phi/2-8A-2B}(\partial\gamma)^2
\\
&+\frac32e^{-\phi-6A-2B}(\partial\chi)^2
+72 e^{-\phi-6A}\omega^2\chi^2
+\frac{1}{4}e^{\phi/2}
\left(
e^{-8A-8B}\varphi^2-3e^{-8A}( m\chi^2+\gamma)^2
\right) 
~.}}
The $F$-form equation of motion is automatically satisfied. The $H$-form equation reduces to,
\eq{\spl{\label{hfeom}
0&= -\nabla^{\mu}\left(
e^{-\phi+4A+2B}\partial_{\mu}\chi
\right) +48\omega^2 e^{-\phi+4A+4B}\chi
+ e^{3\phi/2+6A+4B}m^2\chi
\\
&+2me^{\phi/2+2A+4B}( m\chi^2+\gamma)\chi
-  \varphi
( m\chi^2+\gamma)
~.}}
The $G$-form equation of motion reduces to,
\eq{\label{gfeom1}
0= \nabla^{\mu}\left(
e^{\phi/2+2A+2B}\partial_{\mu}\gamma
\right) -48\omega^2 e^{\phi/2+2A+4B}( m\chi^2+\gamma)+48\omega^2\chi\varphi ~,}
together with the following constraint,
\eq{\label{gfeom2}
0=\frac13 \d\left(
e^{\phi/2+2A-4B}\varphi
\right)
+  (m\chi^2+\gamma)\d\chi+\chi\d\gamma
~.}
The latter can be readily integrated to solve for $\varphi$ in terms of the remaining fields,
\eq{\label{fg}
\varphi=\left(C-m\chi^3-3\gamma\chi\right)e^{-2A+4B-\phi/2}
~,}
where $C$ is an arbitrary constant.

\vfill\break

{\it The Lagrangian}

As we can see from (\ref{tdma}) the scalar $B(x)$ can be reabsorbed in the definition of the 4d metric. We have kept it arbitrary so far with the idea to use the associated 
freedom in order to obtain a 4d effective theory directly in the Einstein frame. This can be accomplished  by choosing, 
\eq{\label{baeq} B=-4A~.} 
With this choice, and taking into account that $\varphi$ is given in eq.~(\ref{fg}), it is straightforward to check that 
the ten-dimensional equations given in (\ref{et1})-(\ref{gfeom1}) all follow from the 4d effective action,
\eq{ \label{ctr}
S_4=\int\d^4 x\sqrt{g}
\left(R - 24 (\partial A)^2  -\dfrac{1}{2} (\partial \phi)^2  
-\dfrac{3}{2} e^{-4A-\phi}(\partial \chi)^2 - \dfrac{1}{32\omega^2} e^{-6A+\phi/2} (\partial \gamma)^2 - V
\right)
~,
}
where the potential $V$ is given by,
\eq{\spl{\label{pot}
V = -120 \omega^2 e^{-8A} &+ \dfrac{1}{2}m^2 e^{-6A+5\phi/2} +\dfrac{3}{2}m^2\chi^2 e^{-10A+3\phi/2} +72\omega^2\chi^2 e^{-12A-\phi} \\
&+ \dfrac{3}{2}(m\chi^2+\gamma)^2 e^{-14A+\phi/2}
+\frac12 \left(C-m\chi^3-3\gamma\chi\right)^2e^{-18A-\phi/2}~.
}}

\subsection{AdS$_4$ solutions revisited}\label{sec:3ads}

The consistent truncation (\ref{ctr}) captures all of the AdS$_4$ solutions of \cite{Lust:2008zd} reviewed in section \ref{sec:ads}. Indeed upon setting the warp factor and the 
dilaton to zero, $A=\phi=0$, and the remaining fields $\gamma$, $\chi$ to constant values, 
imposing the equations of motion amounts to finding a minimum of the potential $V$ of (\ref{pot}). We thus obtain the following three classes of solutions: 

{\it First class}
\eq{H=\pm m\mathrm{Re}\Omega~;~~~
F=\pm\frac{1}{\sqrt{3}} m J~;~~~G=\mp\sqrt{3}m~\!\mathrm{vol}_4-\frac12 m J^2~;~~~ \Omega=-\frac23 m^2
~,}
where it is understood that the sign of $F$ is correlated with that of the external part of $G$, while the sign of $H$ is arbitrary. This can be written equivalently,
{
\eq{\omega=\pm\frac{1}{2\sqrt{3}}m ~;~~~
\chi=\pm\frac{1}{\sqrt{3}} ~;~~~ \gamma= -\frac{4}{3}m ~;~~~ C=\mp \frac{20}{3\sqrt{3}}m
~.}

{\it Second class}
\eq{H=0~;~~~
F=0~;~~~G=\pm\sqrt{5}m~\!\mathrm{vol}_4 ~;~~~\Omega=-\frac12 m^2
~.}
Or, equivalently,
{
\eq{\omega=\pm\frac{1}{2\sqrt{5}}m ~;~~~
\chi=0 ~;~~~ \gamma= 0 ~;~~~ C=\pm \sqrt{5}m
~.}

{\it Third class}
\eq{H=\pm\frac{2}{5}m \mathrm{Re}\Omega~;~~~
F=\pm \frac{1}{\sqrt{15}}m J~;~~~G=\pm \sqrt{\frac{27}{5}}m~\!\mathrm{vol}_4+\frac{3}{10} m J^2~;~~~ \Omega=-\frac{16}{25} m^2
~,}
where it is understood that  the sign of $F$ is correlated with that of the external part of $G$, while the sign of $H$ is arbitrary. Equivalently,
{
\eq{\omega=\pm\frac{1}{\sqrt{15}}m ~;~~~
\chi=\pm\frac{1}{\sqrt{15}} ~;~~~ \gamma= \frac{8}{15}m ~;~~~ C=\pm \frac{32}{3\sqrt{15}}m
~.}
In the above we have noted that $\Omega$ is given by the value of $V/6$ at the minimum, as follows from (\ref{ctr}), (\ref{ei43}). These coincide with the three classes of solutions 
presented in section 11.4 of \cite{Lust:2008zd}, with the third class being the supersymmetric one, cf.~(\ref{sf}).

\subsection{Consistent truncation with condensates}\label{sec:3dil}

In the presence of condensates, the internal $(m,n)$-components of the Einstein 
equations (\ref{et1}) get modified as follows,
\eq{\spl{\label{et1c}
0&=e^{-8A-2B}\nabla^{\mu}\left(
e^{8A+2B}\partial_{\mu}A
\right)+\frac{1}{16}m^2e^{5\phi/2+2A+2B}
+\frac{5}{16}e^{3\phi/2-2A+2B}m^2\chi^2\\
&+\frac18e^{-\phi-4A}(\partial\chi)^2
+18 e^{-\phi-4A+2B}\omega^2\chi^2
+\frac{1}{16}e^{\phi/2}
\left(
3e^{-6A-6B}\varphi^2+7e^{-6A+2B}( m\chi^2+\gamma)^2
\right)\\
&+\frac{1}{256\omega^2}e^{\phi/2-6A}(\partial\gamma)^2
-20e^{2B}\omega^2
-\frac74 e^{A+2B}\text{Im}(\mathcal{B}\omega+\mathcal{C}\omega^*)
\\
&-\frac{1}{32}e^{2A+2B}\left(
21e^{5\phi/4}m\text{Re}\mathcal{A}-\frac{3}{128}(\bar{\Lambda}\Lambda)^2
\right)
+\frac58e^{3\phi/4+2B}m\chi\text{Im}\mathcal{A}\\
&+\frac{3}{32}e^{\phi/4-2A-2B}\varphi\text{Im}\mathcal{A}
-\frac{15}{32}e^{\phi/4-2A+2B}(m\chi^2+\gamma)\text{Re}\mathcal{A}~,}}
where we have taken (\ref{ei43}) into account. The external $(\mu,\nu)$-components read,
\eq{\spl{\label{et2c}
R^{(4)}_{\mu\nu}&=
g_{\mu\nu}\left(\nabla^{2}A+\nabla^{2} B+
8(\partial A)^2+2(\partial B)^2+10\partial A\cdot \partial B\right)
\\
&-8\partial_{\mu}A\partial_{\nu}A-2\partial_{\mu}B\partial_{\nu}B
-16\partial_{(\mu}A\partial_{\nu)}B+8\nabla_{\mu}\partial_{\nu}A+2\nabla_{\mu}\partial_{\nu}B\\
&+\frac32e^{-\phi-4A} 
\partial_{\mu}\chi\partial_{\nu}\chi
+\frac12\partial_{\mu}\phi\partial_{\nu}\phi
+\frac{1}{32\omega^2}e^{\phi/2-6A}\partial_{\mu}\gamma\partial_{\nu}\gamma
 \\
&+\frac{1}{16}g_{\mu\nu}\Big[  -6e^{-\phi-4A}(\partial\chi)^2
-\frac{3}{16\omega^2}e^{\phi/2-6A}(\partial\gamma)^2
\\
&+m^2e^{5\phi/2+2A+2B}
-3m^2\chi^2 e^{3\phi/2-2A+2B} 
\\
&-288e^{-\phi-4A+2B}\omega^2\chi^2-5e^{\phi/2-6A-6B}\varphi^2 
-9e^{\phi/2-6A+2B}( m\chi^2+\gamma)^2 
\\
&
-\frac{1}{2}e^{2A+2B}\left(
21e^{5\phi/4}m\text{Re}\mathcal{A}-\frac{3}{128}(\bar{\Lambda}\Lambda)^2
\right)
-12 e^{A+2B}\text{Im}(\mathcal{B}\omega+\mathcal{C}\omega^*)
\\
&-\frac{9}{2}e^{\phi/4-2A-2B}\varphi\text{Im}\mathcal{A}
+\frac{9}{2}e^{\phi/4-2A+2B}(m\chi^2+\gamma)\text{Re}\mathcal{A}
\Big]
~,}}
while the mixed $(\mu,m)$-components are automatically satisfied. 
The dilaton equation reads,
\eq{\spl{\label{et3c}
0&=e^{-10A-4B}\nabla^{\mu}\left(
e^{8A+2B}\partial_{\mu}\phi
\right)
+\frac32e^{-\phi-6A-2B}(\partial\chi)^2
-\frac{1}{64\omega^2}e^{\phi/2-8A-2B}(\partial\gamma)^2
\\
&-\frac{5}{4}m^2e^{5\phi/2}
-\frac{9}{4}e^{3\phi/2-4A}m^2\chi^2
+72 e^{-\phi-6A}\omega^2\chi^2
\\
&
+\frac{1}{4}e^{\phi/2}
\left(
e^{-8A-8B}\varphi^2-3e^{-8A}( m\chi^2+\gamma)^2
\right) 
+\frac{105}{16}e^{5\phi/4}\text{Re}\mathcal{A}
\\
&
-\frac{45}{16}e^{3\phi/4-2A}m\chi\text{Im}\mathcal{A}
+\frac{3}{16}e^{\phi/4-4A-4B}\varphi\text{Im}\mathcal{A}
+\frac{9}{16}e^{\phi/4-4A}(m\chi^2+\gamma)\text{Re}\mathcal{A}
~.}}
The $F$-form equation of motion is automatically satisfied. The $H$-form equation \eqref{hfeom} is unchanged. 
The $G$-form equation of motion reads,
\eq{\label{gfeom1c}
0= \nabla^{\mu}\left(
e^{\phi/2+2A+2B}\partial_{\mu}\gamma
\right) -48\omega^2 e^{\phi/2+2A+4B}( m\chi^2+\gamma)+48\omega^2\chi\varphi 
+36\omega^2e^{\phi/4+6A+4B}\text{Re}\mathcal{A}
~,}
together with the following constraint,
\eq{\label{gfeom2c}
0=\d\left(
\frac13\varphi e^{\phi/2+2A-4B}
+\frac14e^{\phi/4+6A} \text{Im}\mathcal{A}
\right)
+  (m\chi^2+\gamma)\d\chi+\chi\d\gamma
~.}
The latter can be readily integrated to solve for $\varphi$ in terms of the remaining fields,
\eq{\label{fgc}
\varphi=\left(C-m\chi^3-3\gamma\chi\right) e^{-\phi/2-2A+4B}
-\frac34e^{-\phi/4+4A+4B} \text{Im}\mathcal{A}
~,}
where $C$ is an arbitrary constant.

Upon imposing \eqref{baeq} as before, 
a tedious but straightforward calculation then shows that all the above equations of motion can be obtained from the 
following four-dimensional action,
\boxedeq{ \label{ctrcs}
S_4=\int\d^4 x\sqrt{g}
\left(R - 24 (\partial A)^2  -\dfrac{1}{2} (\partial \phi)^2  
-\dfrac{3}{2} e^{-4A-\phi}(\partial \chi)^2 - \dfrac{1}{32\omega^2} e^{-6A+\phi/2} (\partial \gamma)^2 - V
\right)
~.
}
The action has exactly the same kinetic terms as before, cf.~\eqref{ctr}, but the potential now reads,
\eq{\spl{\label{potcs}
V &= -120 \omega^2 e^{-8A} + \dfrac{1}{2}m^2 e^{-6A+5\phi/2} +\dfrac{3}{2}m^2\chi^2 e^{-10A+3\phi/2} +72\omega^2\chi^2 e^{-12A-\phi} \\
&+ \dfrac{3}{2}(m\chi^2+\gamma)^2 e^{-14A+\phi/2}
+\frac12\varphi^2 e^{18A+\phi/2}-12 e^{-7A} \text{Im}(\mathcal{B}\omega+\mathcal{C}\omega^*)
\\
&+\frac{15}{4}m\chi e^{3\phi/4-8A} \text{Im}\mathcal{A} -\frac{21}{4}e^{5\phi/4-6A} m \text{Re}\mathcal{A}
-\frac{9}{4}e^{\phi/4-10A}(m\chi^2+\gamma) \text{Re}\mathcal{A}\\&+\frac{3}{512}e^{-6A}(\bar{\Lambda}\Lambda)^2~,
}}
where $\varphi$ is non-dynamical and is given by, 
\eq{\label{fgcgg}
\varphi=\left(C-m\chi^3-3\gamma\chi\right) e^{-\phi/2-18A}
-\dfrac34e^{-\phi/4-12A} \text{Im}\mathcal{A}
~.}
It can also be seen that this consistent truncation contains the $S^{1,3}\times M_6$ solutions of section \ref{sec:condensate} as special cases.

\subsection{The Calabi-Yau limit}\label{sec:cylt}

It can be seen from the equations of motion that the limit $\omega\rightarrow0$ can be taken consistently, provided that we first rewrite,
\eq{\label{gx}
\gamma=\gamma_0+ 4\omega\xi~,}
where $\gamma_0$ is constant while $\xi$ is dynamical. 
This  corresponds to  the CY limit, 
in the sense of the vanishing of all $SU(3)$ torsion classes.\footnote{This is more general than the 
usual definition of a CY, as it allows for manifolds with nonvanishing 
fundamental group such as $T^6$.}  

More explicitly, in this case our ansatz for the forms becomes,
\eq{\label{foranscy}
F=m\chi J~;~~~ H=\d\chi \swed J~;~~~
G=\varphi\text{vol}_4+\frac12 (m\chi^2 +\gamma_0)J\swed J-\frac{1}{2}\d\xi\swed\text{Im}\Omega
~,}
and can be seen to automatically satisfy the BI's (\ref{bi}), taking into account that $\d J=\d\Omega=0$.  
All remaining equations of motion can be obtained from those of section \ref{sec:3dil} by first replacing $\gamma$ 
using \eqref{gx} and 
then taking the  $\omega\rightarrow0$ limit. Note that this rewriting allows to keep the dynamical field $\xi$ in the limit.

Moreover  it can be seen that all equations of motion can be integrated into the following Lagrangian,
\eq{ \label{ctrcy}
S^{\text{CY}} =\int\d^4 x\sqrt{g}
\left(R - 24 (\partial A)^2  -\dfrac{1}{2} (\partial \phi)^2  
-\dfrac{3}{2} e^{-4A-\phi}(\partial \chi)^2 - \dfrac{1}{2} e^{-6A+\phi/2} (\partial \xi)^2 - V^{\text{CY}}
\right)
~.
}
The potential $V^{\text{CY}}$ above is given by,
\eq{\spl{\label{potcy}
V^{\text{CY}} =  &+ \dfrac{1}{2}m^2 e^{-6A+5\phi/2} +\dfrac{3}{2}m^2\chi^2 e^{-10A+3\phi/2}+ \dfrac{3}{2}(m\chi^2+\gamma_0)^2 e^{-14A+\phi/2} \\
&
+\frac12\varphi^2 e^{18A+\phi/2}
+\frac{15}{4}m\chi e^{3\phi/4-8A} \text{Im}\mathcal{A} -\frac{21}{4}e^{5\phi/4-6A} m \text{Re}\mathcal{A}
\\
&-\frac{9}{4}e^{\phi/4-10A}(m\chi^2+\gamma_0) \text{Re}\mathcal{A}+\frac{3}{512}e^{-6A}(\bar{\Lambda}\Lambda)^2~,
}}
where,
\eq{\label{fgc}
\varphi=\left(C-m\chi^3-3\gamma_0\chi\right) e^{-\phi/2-18A}
-\frac34e^{-\phi/4-12A} \text{Im}\mathcal{A}
~.}
It can be seen that,  in the absence of condensates, 
unless all flux is zero, the potential is non-negative and only has runaway minima. However this need no longer be the case in the presence of nonvanishing condensates.

\vfill\break

{\bf Massless limit}

A further truncation to two scalars, the warp factor $A$ and the dilaton $\phi$, 
can be obtained by taking the massless limit, $m=0$, while at the same time setting  $\chi$, $\gamma=0$. This amounts to the following flux ansatz:
\eq{\label{foranscym}
F=0~;~~~ H=0~;~~~
G=\varphi\text{vol}_4
~,}
which is of Freund-Rubin type, and automatically satisfies the BI's (\ref{bi}). Moreover the remaining form equations reduce to a single constraint,
\eq{\label{fgm}
\varphi=C~\!e^{-\phi/2-18A}
-\frac34e^{-\phi/4-12A} \text{Im}\mathcal{A}
~,}
where $C$ is an arbitrary constant. It can then be seen that all equations of motion can be integrated to the following Lagrangian,
\eq{ \label{ctrcym}
S^{\text{CY}}_{0} =\int\d^4 x\sqrt{g}
\left(R - 24 (\partial A)^2  -\dfrac{1}{2} (\partial \phi)^2  
-V_0^{\text{CY}}
\right)
~,
}
where the potential $V_0^{\text{CY}}$ is given by,
\eq{\label{potcyml}
V_0^{\text{CY}} =  
\frac12\varphi^2 e^{18A+\phi/2}+\frac{3}{512}e^{-6A}(\bar{\Lambda}\Lambda)^2~.
}}

\section{Discussion}\label{sec:disc}

We have constructed a 
consistent truncation of (massive) ten-dimensional IIA supergravity on NK or CY spaces in the presence of dilatino condensates.  
 It has been argued in \cite{Cvetic:2000dm} that, at least in the case of sphere reductions, 
 the existence of a consistent  truncation of a higher-dimensional supersymmetric theory to the bosonic sector of a supersymmetric 
lower-dimensional theory implies the existence of  a consistent  truncation to the full lower-dimensional theory.   
In  the limit of vanishing condensates, it would be interesting to try to establish a dictionary between the present paper and the bosonic sector of the $\mathcal{N}=2$ four-dimensional truncation considered in \cite{KashaniPoor:2007tr}.

The solutions  of the present paper are formal, in that 
we have not offered  any concrete mechanism for the generation of the dilatino condensate. 
More importantly perhaps, one would need to provide a controlled setting in which the dilatino condensate is not 
negligible compared to other quantum corrections. One place  where one might be able to 
carry out this calculation would be in the context of the the uncompactified theory: since the two-derivative low-energy effective 
action of the ten-dimensional IIA superstring is expected to be exact to all orders in string perturbation, a potential fermion condensate generated nonperturbatively in 
the string coupling (e.g.~via gravitational instantons as in \cite{Konishi:1988mb}) would not have to ``compete'' with any string-loop corrections. We hope to return to this in the future.

\vfill\break

\section*{Acknowledgment}

We are grateful to Derek Harland for bringing reference \cite{Foscolo:2015vqa} to our attention.

\appendix

\section{SU(3)-structure conventions}\label{app:a}

We follow the conventions of \cite{Lust:2004ig} for the SU(3) structure. Note in particular that we are using superspace conventions for the forms.

Let $\eta$ be the SU(3)-singlet spinor of $M_6$. 
The intrinsic torsion parametrizes the 
failure of the spinor $\eta$ to be covariantly constant. It  decomposes into five modules (torsion classes) 
${W}_1, \dots, {W}_5$: 
\eq{\spl{
\nabla_m\eta&=\frac{1}{2}\left(W_{4m}^{(1,0)}+W_{5m}-\mathrm{c.c} \right)\eta\\
&+\frac{1}{16}\left(
4W_1g_{mn}-2W_4^p\Omega_{pmn}+4iW_{2mn}-iW_{3mpq}\Omega^{pq}{}_n
\right)\gamma^n\eta^c ~,
\label{torsionclasses}
}}
where $W_1$ is a complex scalar, $W_2$ is a complex (1,1)-traceless form, $W_3$ is a real traceless $(2,1)+(1,2)$ form, $W_4$ is a real one-form and $W_5$ is a (1,0) form. 
In terms of $SU(3)$ representations,
\eq{
W_1\sim {\bf 1}\oplus  {\bf {1}};~~~~~ W_2\sim {\bf 8}\oplus  {\bf {8}}; 
~~~~~ W_3\sim {\bf 6}\oplus  {\bf \bar{6}}; 
~~~~~ W_4\sim {\bf 3}\oplus  {\bf \bar{3}};  ~~~~~ W_5\sim {\bf {3}}~.
}
Equivalently these torsion classes parameterize the failure of closure of the exterior derivatives of $J$, $\Omega$. Explicitly,
\eq{J=i\tilde{\eta}\ga_{(2)}\eta^c~;~~~\Omega=\tilde{\eta}\ga_{(3)}\eta
~,}
and,
\eq{\spl{
dJ&=\frac{3}{2}\mathrm{Im}(W_1^*\Omega)+W_4\wedge J+W_3\;,\\
d\Omega&= W_1 J\wedge J+W_2 \wedge J+ W_5^* \wedge \Omega~.
\label{torsionclassesb}
}}
The following identities are useful in the decomposition of the fermionic equations 
of motion,
\eq{\spl{
0&=(\Pi^+)_m{}^n\gamma_n\eta^c\\
\gamma_{mn}\eta&=iJ_{mn}\eta+\frac{1}{2}\Omega_{mnp}\gamma^p\eta^c\\
\gamma_{mnp}\eta^c&=-3iJ_{[mn} \gamma_{p]}\eta^c-\Omega^*_{mnp}\eta~.
}}

{\bf Nearly K\"{a}hler}

For a NK manifold the previous formulas simplify as follows. 
The spinor derivative reads,
\eq{ \nabla_m\eta=i\omega\gamma_m\eta^c ~,
\label{torsionclassesnk}
}
where $\omega:=-iW_1/4$ and $\omega$ is a real constant.  
Equivalently we have,
\eq{\spl{
\d J&=-6\omega\mathrm{Re}\Omega \\
\d\mathrm{Im}\Omega&= 4\omega J\wedge J ~.
\label{torsionclassesbnk}
}}

\section{Spinor conventions}\label{sec:conventions}

Our spinor conventions are identical to those in  \cite{Lust:2004ig}, cf.~appendix A therein, except for a slight change of notation with respect to the 
$SU(3)$-singlet spinor of the internal manifold $M_6$, which is denoted $\eta$ here and corresponds to $\eta_+$ of \cite{Lust:2004ig}. 
Moreover the $\eta_-$ of that reference corresponds to $\eta^c:=C\eta^*$ here.

We note the following useful properties of spinor bilinears in six dimensions,
\eq{\spl{
(\widetilde{\psi}_{\pm}\ga_{(2p)}\chi_{\pm})&=(\widetilde{\psi}_{\pm}\ga_{(2p+1)}\chi_{\mp})=0\\
(\widetilde{\psi}\ga_{(p)}\chi)&=(-1)^{\frac12 p(p+1)}(\widetilde{\chi}\ga_{(p)}\psi)
~,}}
where $\psi_{+}$, $\chi_{+}$ are arbitrary commuting Weyl spinors of positive chirality; $\psi_{-}$, $\chi_{-}$ are arbitrary commuting Weyl spinors of negative chirality; 
$\psi$, $\chi$ are arbitrary commuting Dirac spinors.

Spinor bilinears in four dimensions obey,
\eq{\spl{
(\widetilde{\psi}_{\pm}\ga_{(2p+1)}\chi_{\pm})&=(\widetilde{\psi}_{\pm}\ga_{(2p)}\chi_{\mp})=0\\
(\widetilde{\psi}\ga_{(p)}\chi)&=(-1)^{\frac12 p(p-1)}(\widetilde{\chi}\ga_{(p)}\psi)
~,}}
where $\psi_{+}$, $\chi_{+}$ are now arbitrary {anti}-commuting Weyl spinors of positive chirality; $\psi_{-}$, $\chi_{-}$ are arbitrary anti-commuting Weyl spinors of negative chirality; 
$\psi$, $\chi$ are arbitrary anti-commuting Dirac spinors.

%
%


\end{document}